\documentclass[twocolumn,aps,showpacs,preprintnumbers,amsmath,amssymb]{revtex4}

\usepackage{bm}
\newcommand{\goa}{\hspace{2pt}^{>}\hspace{-7pt}_{\sim}}
\newcommand{\loa}{\hspace{2pt}^{<}\hspace{-7pt}_{\sim}}

\begin{document}


\title{On the stable configuration of ultra-relativistic material
 spheres.\\ The solution for the extremely hot gas}
\author{Lubo\v{s} Neslu\v{s}an}%
 \email{ne@ta3.sk}
\affiliation{%
 Astronomical Institute of the Slovak Academy of Sciences,\\
 05960 Tatransk\'{a} Lomnica, Slovakia
}%

\date{\today}

\begin{abstract}
   During the last stage of collapse of a compact object into the
horizon of events, the potential energy of its surface layer decreases
to a negative value below all limits. The energy-conservation law
requires an appearance of a positive-valued energy to balance the
decrease. We derive the internal-state properties of the ideal gas
situated in an extremely strong, ultra-relativistic gravitational
field and suggest to apply our result to a compact object with the
radius which is slightly larger than or equal to the Schwarzschild's
gravitational radius. On the surface of the object, we find that the
extreme attractivity of the gravity is accompanied with an extremely
high internal, heat energy. This internal energy implies a
correspondingly high pressure, the gradient of which has such a
behavior that it can compete with the gravity. In a more detail,
we find the equation of state in the case when the magnitude of the
potential-type energy of constituting gas particles is much larger
than their rest energy. This equation appears to be identical with
the general-relativity condition of the equilibrium between the
gravity and pressure gradient. The consequences of the identity are
discussed.
\end{abstract}

\pacs{03.75.Hh, 05.70.Ce, 51.30.+i, 52.27.Ny}
\keywords{thermodynamical functions -- dense matter -- equation of state
 -- gravitation -- relativity}

\maketitle


\begin{quote}
{\em Motto:} "Chandrasekhar shows that a star of mass greater than a
certain limit remains a perfect gas and can never cool down. The star
has to go on radiating and radiating and contracting and contracting
until, I suppose, it gets down to a few kilometers' radius when gravity
becomes strong enough to hold the radiation and the star can at least
find peace. I felt driven to the conclusion that this was almost a
{\em reductio ad absurdum} of the relativistic degeneracy formula.
Various accidents may intervene to save the star, but I want more
protection than that. I think that there should be a law of Nature to
prevent the star from behaving in this absurd way."\\
\indent \hfill{Sir Arthur Eddington \cite{chandra72}}
\end{quote}

\section{Introduction}\label{sec1}

   After a star with a mass lower than the Oppenheimer-Volkoff limit has
spent its nuclear fuel, it becomes either the white dwarf or neutron
star. The rest of its thermal energy is irradiated into the neighbouring
space and, thus, its temperature permanently decreases, down to a value
approaching the absolute zero. Its internal structure can be described
with the help of theory of cold degenerated electron or neutron gas.

   In this work, we polemize with the current theory of the
supplementary case: the end-state of the star with the final mass
exceeding the Oppenheimer-Volkoff limit. Namely, there are several
serious arguments that such a star cannot end as a cold object. Even if
it was cooled enough during the first stage of the collapse, its internal
energy and temperature can be expected to increase over all limits, when
its radius approaches the horizon of events, which was found at the
Schwarzschild gravitational radius, $R_{g}$. More specifically, the
increase of internal energy can be expected when the proper radius of
the object is reduced below, say, $1.01\, R_{g}$ or $1.001\, R_{g}$.

   Our considerations are based on the classical quantum statistics and
classical general relativity (with no-hair theorems valid), which both
are well-known a long time. Only new element concerning the assumptions
in this work is a term corresponding with a potential-type energy in
the formulas of quantum-statistics equations of state for the extremely
hot gas. In fact, our idea is not new, in principle. The
potential-energy term is well known in the classical Maxwell-Boltzmann
impulse-distribution law of ideal gases having a low internal energy.
In the latter, it is identical to the classical potential energy and
situated in the Boltzmann part of the distribution law.

   In our description of gas properties, we use the formula for the
relativistic energy which is suitable in a situation when the rest
energy of gas particles can be neglected. In the classical works by
Landau \cite{landau32}, Chandrasekhar \cite{chandra35}, or Oppenheimer
\& Volkoff \cite{oppenheimer:volkoff39}, the influence of conservative
gravitational field on the state properties of the gas was not taken
into account. A step toward an inclusion of this influence occurred in
more recent studies of stellar collapse and neutron stars. Prakash et
al. \cite{prakash:etal97} considered the potential-energy term in
course to provide the equation of state of a hot gas, which constitutes
a neutron star a short time after its formation. The authors, however,
considered that approximative form of the relativistic formula for
energy, which is applicable at the values much lower than the rest
energy. Their work was focused especially on the nuclear reactions
and alternative forms of matter constituting the star. They did not
deal with any collapse going on down to the gravitational radius. Very
recently, Moustakidis \& Panos \cite{moustakidis:panos09}, who
continued in the work of Prakash et al., provided an equation of state
for a hot nuclear matter with the $\beta$-decay equilibrium. This kind
of matter was again assumed to constitute a neutron star in less extreme
conditions, allowing neutrons to be unstable, than we are going to
consider.\\

   The following arguments can be presented to support the concept of
the occurrence of extremely hot gas in an object with radius approaching
$R_{g}$.

   (1) Let us consider a radiation coming from an ultra-relativistic
compact object to an observer at a relatively large distance. In his own
reference frame, the observer detects the radiation which is
red-shifted. In other words, its energy is lower in the reference
frame of the observer than was at the moment of its emission, in the
reference frame of the source. In a reverse situation, when a radiation
comes onto the surface of the compact objects from a distant source, its
wave-length is blue-shifted. This time, its energy is increasing with
time. It is higher in the object-surface reference frame than in the
frame of the distant source.

   The numerical calculation reveals that the frequency of the radiation
coming to the horizon at $R_{g}$ from an outer space increases over all
limits implying the corresponding energy to increase over all limits.
Such the increase must, however, occur not only in the case of
radiation, but in the case of non-zero rest-mass particles, when these
are falling down, as well. The requirement of the local conservation of
energy in a volume of the compact-object-surface layer means that the
increase must be, on the other hand-side, balanced by a decrease of that
type energy of the surface layer, which has a character of potential
energy. We refer to this energy as the "potential-type energy".

   Therefore, when the collapsing-object radius approaches the
gravitational radius, its internal thermal energy must increase over all
limits, because we do not know any efficient mechanism of the
extreme-rate cooling. A fraction of internal energy is irradiated in
form of photons and carried out by neutrinos only from the object's
surface. Even in this case, the efficiency of both cooling processes
decreases down to zero when the gravitational radius is approached (see
our more specific considerations in Sect.~\ref{sec6}). Thus, an extreme
accumulation of the internal energy seems to be inevitable in the last
stage of the collapse to the horizon and can be the cause of a strong
pressure with a steep gradient resisting to the gravity. After the
object's radius reaches the horizon, the accumulated energy is trapped
inside the object forever. (Some exceptions resulting from our work
are discussed in Sect.~\ref{sec7}. In the context of our work, we do not
consider fine effects as, e.g., Hawking's evaporation of black holes.)

   (2) Tolman \cite{tolman30b} and Tolman \& Ehrenfest
\cite{tolman:ehrenfest30} found that inside the sphere filled in with an
ideal material fluid, the product of both proper temperature of the
fluid, $T$, and square root of component $g_{44}$ of metrique tensor is
a finite constant, i.e.
\begin{equation}\label{equivT}
T\sqrt{g_{44}} = T\exp(\nu/2) = T_{c} = const.
\end{equation}
($\exp(\nu )$ is an alternative expression of $g_{44}$; see
Eq.(\ref{line_elem}) and the text below this equation). It means that
if a volume of the fluid is situated near the surface of the sphere,
the radius of which approaches $R_{g}$, then $g_{44} \rightarrow 0$ and
the temperature of the fluid must approach infinity the product of both
could be the finite constant.

   (3) At the same time, Tolman \& Ehrenfest \cite{tolman:ehrenfest30}
demonstrated that the quantity $P_{rad}\, g_{44}^{2}$ is equal to a
finite constant inside the sphere filled in with the black-body
radiation. $P_{rad}$ is the radiation pressure. When a volume of
gaseous material fluid sinks to a more and more curved space-time at
the gravitational radius, radiation pressure $P_{rad}$ must be
magnified over all limits to keep the constancy of
$P_{rad}\, g_{44}^{2}$. The requirement of this finite constancy also
supports the hot concept of final stage of very massive stars.\\

   Considering all these facts, we search for the "hot" solution of
the static configuration of a spherically symmetric object with no
generated energy. Specifically, we give a brief summary of the theory
and results that have been achieved up to date in Sect.~\ref{sec2}. The
introduction of a new modification of concerning fundamental theory
with some obvious steps toward the solution of established fundamental
integrals is presented in Sect.~\ref{sec3}. In Sect.~\ref{sec4}, we
derive the solution, which is applicable on an extremely hot gas. In
more detail, we derive the state quantities of the ideal gas situated
in a region of extremely curved space-time of the stable spherically
symmetric object, whereby the curvature is so large that the magnitude
of the potential-type energy of a given particle is much greater than
its rest energy. In such a gas, a radiation must necessarily occur. So,
an influence of the radiation pressure on the object's stability is
discussed in Sect.~\ref{sec5}. In Sect.~\ref{sec6}, we attempt roughly
sketch a behavior of the collapse of object before it disappears
below horizon. Finally, we summarize our most important results and
discuss some implications toward a potential observational evidence of
this kind of objects, in Sect.~\ref{sec7}.

\section{The old theory}\label{sec2}

   The theory of the object contracting to approach a zero radius was
completed by Oppenheimer \& Volkoff \cite{oppenheimer:volkoff39}. They
went on in the work started by Landau \cite{landau32} who showed that
for a model of the star consisting of a cold degenerate Fermi-Dirac gas,
after all the elements inside available for thermonuclear reactions are
used up, there exist no stable equilibrium configurations for masses
greater than a certain critical mass. In contrast to Landau, who
considered the Newtonian gravitational theory, Oppenheimer \& Volkoff
accounted for the general relativistic effects, which were expected for
the very high masses and densities.

   Oppenheimer \& Volkoff \cite{oppenheimer:volkoff39} considered the
static line element known for the spherical symmetry in the form
\begin{equation}\label{line_elem}
ds^{2} = -e^{\lambda}dr^{2} - r^{2}d\theta^{2} -
 r^{2}\sin^{2}\theta\, d\phi^{2} + e^{\nu}dt^{2},
\end{equation}
where $r$, $\theta$, and $\phi$ are common spherical coordinates, $t$ is
time, and $\lambda$, $\nu$ are functions of $r$. They further considered
the Einstein equations of field written for the case of spherical
symmetry \cite{eddington23}, in which they put the expressions of the
non-zero components of energy tensor $T_{1}^{1} = T_{2}^{2} = T_{3}^{3}
= -P_{gas}$ and $T_{4}^{4} = E_{gas}$ expressed in this form by Tolman
\cite{tolman30a} (see Eqs. (\ref{Einstein1}), (\ref{Einstein2}), and
(\ref{Einstein3}), in Sect.~3). $P_{gas}$ and $E_{gas}$ are the gas
pressure and its energy density (internal energy of gas per unit
volume), respectively. The contra-variant components of the energy
tensor for an ideal material fluid, $T^{11} = T^{22} = T^{33} = P_{gas}$
and $T^{44} = E_{gas}$, were specified earlier by Eddington
\cite{eddington23}.

   To solve the field equations, Oppenheimer \& Volkoff established the
auxiliary function $u = u(r)$ equal to
\begin{equation}\label{u}
u = \frac{1}{2}r(1 - e^{-\lambda}),
\end{equation}
with the help of which the equations of field could be written in form
(in SI units used throughout our paper)
\begin{equation}\label{OV1}
\frac{du}{dr} = \frac{4\pi G}{c^{4}}E_{gas}r^{2},
\end{equation}
\begin{equation}\label{OV2}
\frac{dP_{gas}}{dr} = -\frac{P_{gas} + E_{gas}}{r(r - 2u)}
 \left[ \frac{4\pi G}{c^{4}}P_{gas}r^{3} + u\right] ,
\end{equation}
where $G$ is the gravitational constant and $c$ is the velocity of
light. For $P_{gas}$ and $E_{gas}$, they supplied the parametric form of
the equation of state, for the cold degenerated Fermi-Dirac gas, found
by Chandrasekhar (\cite{chandra35}, appendix). After the analysis of the
possible solutions of final equations, Oppenheimer \& Volkoff
concluded that there was not any stable solution for the object
having the mass exceeding the finite limit they specified.

   If their conclusion was indisputable, our re-opening the problem on
the stable configuration would be meaningless. However, we find three
of their assumptions disputable. At first, they assumed that the
collapsing object is cold enough for the theory of the cold degenerated
Fermi-Dirac-statistics gas to be applicable. As we argued in
Sect.~\ref{sec1}, the decreasing potential-type energy should be
balanced by an increase of the thermal energy and this process leads,
instead, to a hot gas. The extreme increase occurs especially at the
last stage of collapse in proper time, but above the event horizon, when
the physical radius of collapsing object is close to $R_{g}$ and
metrique-tensor components $g_{11}$  and $g_{44}$ start significantly
deviate from unity. So, even if the object was cold enough at a proper
radius several times larger than $R_{g}$, it must be highly heated
during this last stage.

   Interestingly, Oppenheimer \& Volkoff noted, in their paper
\cite{oppenheimer:volkoff39}, that for those singular solutions in which
$g_{44}$ vanishes, it is conceivable that the temperature may be high
with the respect to the Tolman \& Ehrenfest \cite{tolman:ehrenfest30}
relation $T\sqrt{g_{44}} = const.$ And, the equation of state reduces to
$E_{gas} = 3P_{gas}$ in the limit of $T \rightarrow \infty$ (which we
analyse in Sect.~\ref{sec4}). However, they supposed the appropriate
conditions for the very high temperature only in the center of the
object. They did not noticed that the appropriate conditions may also
exist in the region around the border of the sphere of radius equal to
$R_{g}$.

   At second, they required the pressure approached zero at the surface
of the object, i.e. for $r = R_{b}$ ($R_{b}$ is the proper radius of the
object, i.e. the distance of the object's surface from its center
measured in the object's surface reference frame). In every-day
experience, we know, however, a lot of examples of pressure or density
discontinuities: the border between a rocky earth surface and air, the
sea level and air, etc. Another example can be found in the outer space,
where neutron stars are expected to have a very large density at the
surface, but, of course, zero density above the surface. We cannot
apriori exclude the possibility that the pressure (or energy density or
particle concentration) is finite in the surface layer between radii
$R_{b} - dr$ and $R_{b}$ and zero in the adjacent layer above, between
$R_{b}$ and $R_{b} + dr$.

   At third, they started the numerical integration of the final
equations with the initial value $u_{o} = u(r=0)$ being in the interval
$0 \geq u_{o} \geq -\infty$. They excluded the possibility $u_{o} > 0$
arguing that the component of metrique tensor $g_{11} = -e^{\lambda}$
must never be positive, therefore $u_{o} = 0$ for all finite values of
$e^{-\lambda}$ and $u_{o} \leq 0$ for infinite values of $e^{-\lambda}$
at the origin. The argument of not positive $-e^{-\lambda}$ is, however,
questionable. Let us consider the object which already collapsed below
the gravitational radius. In the interval of distances between $R_{b}$
and $R_{g}$, the well-known outer Schwarzschild solution is applicable.
According to this solution $-e^{\lambda} = -1/(1 - R_{g}/r) > 0$ for
$R_{b} < r < R_{g}$. Therefore, if we rejected the possibility
$-e^{\lambda} > 0$, there would not exist not only any solution for
the stable configuration of an extinct star, but we would not have any
acceptable description of the field for $r < R_{g}$, either.
Consequently, the reason for the infinite collapse due to the
gravitational attraction would disappear. But we believe that it is
reasonable to retain the possibility $-e^{\lambda} > 0$. And, if we
accept this possibility for $r < R_{g}$, then why we should not doing
so for $r \geq R_{g}$?

\section{The basic considerations}\label{sec3}

   Let us consider a material object, which is spherically symmetric
and ultra-compact, i.e. its proper physical radius, $R_{b}$, is only
slightly larger than its Schwarzschild gravitational radius, $R_{g}$.
To better explain the consequences resulting from the new formulas we
find below, we consider the idealized, non-rotating object that is in or
near its supposed stable configuration. Hereinafter, we refer to this
object as the "ultra-relativistic material sphere" (URMS).

   In this section, we summarize the well-known fundamental equations
of the problem of URMS stable configuration and describe several first
steps to obtain the solution for the stable URMS. Within the summary,
we introduce our own modification of the fundamental equations necessary
for taking the strong, ultra-relativistic gravitational field into
account. The steps toward the solution can be made generally for the gas
consisting of a single kind of particles having spin 1/2 (e.g.
neutrons). No other assumptions are needed.

   In the following, we consider the reference frame related to a volume
of URMS' surface layer. If the URMS is stable, then the number density,
$n_{gas}$, pressure, $P_{gas}$, and internal energy per unit volume,
$E_{gas}$, of the gas in the surface layer can generally be calculated,
as the functions of the impulse $p$, with the help of well-known
integrals (e.g.
\cite{hk94})
\begin{equation}\label{ngen}
n_{gas} = \int_{p} n_{gas}(p)4\pi p^{2} dp,
\end{equation}
\begin{equation}\label{Pgen}
P_{gas} = \frac{1}{3}\int_{p} n_{gas}(p) p\mathcal{V} 4\pi p^{2} dp,
\end{equation}
\begin{equation}\label{Egen}
E_{gas} = \int_{p} n_{gas}(p)\mathcal{E} (p) 4\pi p^{2} dp,
\end{equation}
where the kinetic energy $\mathcal{E}$ and velocity $\mathcal{V}$ of
a particle can be calculated by the well-know relativistic formula
\begin{equation}\label{Wk1}
\mathcal{E}(p) = \sqrt{c^{2}p^{2} + W_{o}^{2}} - W_{o},
\end{equation}
\begin{equation}\label{v}
\mathcal{V} = \frac{\partial \mathcal{E}}{\partial p} = \frac{c^{2}p}
 {\sqrt{c^{2}p^{2} + W_{o}^{2}}}.
\end{equation}
Symbol $W_{o}$ stands for the particle's rest energy.

   The number density $n_{gas}(p)$ for the ideal gas is
\begin{equation}\label{nfp}
n_{gas}(p) = K_{N}\sum_{j} \frac{f_{j}} {\exp \{ [W_{j}(p) - \mu ]
 /(kT)\} + 1},
\end{equation}
where $K_{N}$ is a normalization constant, $k$ is the Boltzmann
constant, $T$ is the temperature of the gas, $f_{j}$ is the degeneracy
of state $j$ (i.e., it gives the number of states having the same energy
$W_{j}$), and $\mu$ is the chemical potential.

   The energy state of a particle $W_{j}$ in the argument of exponential
in Eq.(\ref{nfp}) must be of the type varying with impulse $p$ and with
the force field in which the gas is situated. In the limit of a weak
field, the part $W_{j}/(kT)$ of the argument in Eq.(\ref{nfp}) must
converge to expression $p^{2}/(2mkT) + W_{p}/(kT)$ figuring in the
Maxwell-Boltzmann distribution law. So, the energy does not contain the
rest-energy term, $W_{o}$. It is only the sum of kinetic and
potential-type energy, i.e.
\begin{equation}\label{W}
W_{k+p} = \sqrt{c^{2}p^{2} + W_{o}^{2}} - W_{o} - \chi ,
\end{equation}
where $-\chi$ is the potential-type energy of the particle. By the
convention used in our work, the fact that the gravity is attractive
force is expressed with sign minus in front of $\chi$, in Eq.(\ref{W}).
Within this convention, it is always valid that $\chi \geq 0$. 
Completing the description of the energy state with term $-\chi$ is
an essential point of our work. In Sects.~\ref{sec4}$-$\ref{sec7}, we
derive some important consequences of this additional term.

   The relativistic formula (\ref{W}), but with the rest-energy term
included, is well-known in quantum physics. Its validity was proved,
for example, in the construction of the Klein-Gordon equation, which is
the basis to derive the well-known Dirac equations providing the energy
terms of hydrogen atom, which are in a very good agreement with the
experimental values. In this usage, $-\chi$ is identified to the
classical Coulomb energy for a static system. In the strong,
ultra-relativistic gravitational field, the function $\chi = \chi (r)$
cannot be, of course, identified with the well-known formula for the
potential energy in the Newtonian gravitational field. The explicit form
of this function for a strong field is unknown. We discuss its
dependence on the quantity $\nu$ (in metrique tensor; see
Eq.(\ref{line_elem})) in the case of the extremely strong field in the
next section. For now, we profit from working with the implicit $\chi$.
Fortunately, a more specific expression of $\chi$ appears not necessary.

   In the context of function $\chi$, the "Fermi impulse", $p_{F}$,
corresponding to the Fermi energy, is
\begin{equation}\label{pf}
p_{F} = \frac{1}{c}\sqrt{(\chi + W_{o} + \mu )^{2} - W_{o}^{2}}.
\end{equation}
This relation can be derived requiring the argument of exponential in
Eq.(\ref{nfp}) to be zero and identifying the energy $W_{k+p}$ given by
Eq.(\ref{W}) to the Fermi energy of the particle. We note that
$p_{F} = \chi /c$ for $\chi \gg W_{o}$ and, of course,
$p_{F} \rightarrow \infty$ for $\chi \rightarrow \infty$.

   The energy states, $W_{j}$, in Eq.(\ref{nfp}) differ each other by
their rest energies, $W_{oj}$, for various kinds of the particles. Since
we consider, in this section, only the gas consisting of one-half spin
particles of one kind (neutrons), the sum in Eq.(\ref{nfp}) consists of
only a singe term ($j = 1$) and $f_{1} = 2$.

   The size of impulse $p$ can vary from zero to infinity. So, the
number density, pressure, and internal energy per unit volume are
the integrals
\begin{equation}\label{basicn}
n_{gas} = K\int_{0}^{\infty} \frac{p^{2}}{\exp\left(
 \frac{\sqrt{c^{2}p^{2} + W_{o}^{2}} - W_{o} - \chi - \mu}{kT}\right)
 + 1} dp,
\end{equation}
\begin{eqnarray}\label{basicP}
P_{gas} = \frac{Kc^{2}}{3}\int_{0}^{\infty} \frac{p^{4}}
 {\sqrt{c^{2}p^{2} + W_{o}^{2}}}. \nonumber \\
 .\left[ \exp\left( \frac{\sqrt{c^{2}p^{2} + W_{o}^{2}} - W_{o} - \chi
 - \mu}{kT}\right)  + 1 \right]^{-1} dp,
\end{eqnarray}
\begin{eqnarray}\label{basicE}
E_{gas} = K\int_{0}^{\infty} \frac{p^{2}\left( \sqrt{c^{2}p^{2} +
 W_{o}^{2}} - W_{o}\right) dp}{\exp\left( \frac{\sqrt{c^{2}p^{2} +
 W_{o}^{2}} - W_{o} - \chi - \mu}{kT}\right) + 1} = \nonumber \\
 = 3P_{gas} - W_{o}n_{gas} + KW_{o}^{2}\int_{0}^{\infty}\frac{p^{2}}
 {\sqrt{c^{2}p^{2} + W_{o}^{2}}}. \nonumber \\
 .\left[ \exp\left( \frac{\sqrt{c^{2}p^{2} + W_{o}^{2}} - W_{o} -
 \chi - \mu}{kT}\right)  + 1 \right]^{-1} dp,
\end{eqnarray}
respectively. We denoted $K = 8\pi K_{N}$.

   For the URMS, the Einstein's equations of field can be written in
their form for the spherical symmetry \cite{eddington23},
\cite{tolman30a} (see also \cite{tolman69}, \S95):
\begin{equation}\label{Einstein1}
\frac{8\pi G}{c^{4}} P = \exp(-\lambda )\left( \frac{1}{r}\frac{d\nu}{dr}
 + \frac{1}{r^{2}}\right) - \frac{1}{r^{2}},
\end{equation}
\begin{eqnarray}\label{Einstein2}
\frac{8\pi G}{c^{4}} P = \exp(-\lambda )\left[ \frac{1}{2}
 \frac{d^{2}\nu}{dr^{2}} - \frac{1}{4}\frac{d\lambda}{dr}\frac{d\nu}{dr}
 + \right. \nonumber \\
 + \frac{1}{4}\left( \frac{d\nu}{dr}\right) ^{2} + \left. \frac{1}{2r}
 \left( \frac{d\nu}{dr} - \frac{d\lambda}{dr}\right) \right] ,
\end{eqnarray}
\begin{equation}\label{Einstein3}
\frac{8\pi G}{c^{4}} E = \exp(-\lambda )\left( \frac{1}{r}
 \frac{d\lambda}{dr} - \frac{1}{r^{2}}\right) + \frac{1}{r^{2}},
\end{equation}
in the SI units. $P$ and $E$ are the total pressure and total energy
density in a given volume of space in the proper reference frame. The
cosmological constant, $\Lambda$, is put zero, since we assume that
the URMS can occupy a volume many orders of magnitude smaller than
the cosmological distance-scale.

   It is well-known \cite{tolman30a}, \cite{tolman69} that we can
derive the equation
\begin{equation}\label{basicequ}
\frac{dP}{dr} = -\frac{E + P}{2}\frac{d\nu}{dr},
\end{equation}
from Eqs.(\ref{Einstein1})$-$(\ref{Einstein3}). This is the relativistic
analogue of the Newtonian relation
\begin{equation}\label{Newton_cond}
\frac{dP}{dr} = \rho\frac{d\Psi}{dr} = -\rho \frac{GM_{r}}{r^{2}}
\end{equation}
giving the dependence of total-pressure gradient, $dP/dr$, on the
gravitational potential, $\Psi$ (\cite{tolman69}, \S95). This equation is
also known as the equation of the hydrostatic equilibrium in the theory
of internal structure of stars (e.g. \cite{hk94}, chapt.~1). Quantity
$\rho$ is the density of stellar plasma at the distance $r$ from the
stellar center and $M_{r}$ is the mass within radius $r$. There is an
essential difference between the relativistic Eq.(\ref{basicequ}) and
classical-physics Eq.(\ref{Newton_cond}). The latter does not contain
any term corresponding to term $-(P/2)d\nu /dr$ in Eq.(\ref{basicequ}).
Recently, Rappaport et al. \cite{rappaport:etal08} and Schwab et al.
\cite{schwab:etal09} represented this term as the self-gravity of
pressure.

   Our task to find a stable configuration of the URMS will be solved,
when we find a reasonable combination of functions $P$, $E$, and $\nu$,
for which Eq.(\ref{basicequ}) is satisfied. For the gas, we calculate
the corresponding, gas-related part of the left-hand side of
Eq.(\ref{basicequ}) derivating Eq.(\ref{basicP}), in which the
temperature, $T$, and potential-type energy, -$\chi$, are expected to
be the functions of radial distance, $r$, from the center of the URMS.
(Since we assume the mono-particle gas, the chemical potential, $\mu$,
is obviously not any function of $r$.) We obtain
\begin{eqnarray}\label{gradP2}
\frac{dP_{gas}}{dr} = \frac{Kc^{2}}{3kT^{2}}\frac{dT}{dr}I_{P1} +
 \frac{Kc^{2}}{3kT}\left( \frac{d\chi}{dr} - \right. \nonumber \\
 \left. - \frac{W_{o} + \chi + \mu}{T} \frac{dT}{dr}\right) I_{P2},
\end{eqnarray}
where
\begin{eqnarray}\label{IP1}
I_{P1} = \int_{0}^{\infty} \frac{p^{4} \exp \left( \frac{
 \sqrt{c^{2}p^{2} + W_{o}^{2}} - W_{o} - \chi - \mu}{kT}\right) }
 {\left[ \exp \left( \frac{\sqrt{c^{2}p^{2} + W_{o}^{2}} - W_{o} - \chi
 - \mu}{kT}\right) + 1\right] ^{2}} dp
\end{eqnarray}
and
\begin{eqnarray}\label{IP2}
I_{P2} = \int_{0}^{\infty} \frac{p^{4} \exp \left( \frac{
 \sqrt{c^{2}p^{2} + W_{o}^{2}} - W_{o} - \chi - \mu}{kT}\right) }
 {\sqrt{ c^{2}p^{2} + W_{o}^{2}}}. \nonumber \\
 .\left[ \exp \left( \frac{\sqrt{c^{2}p^{2} + W_{o}^{2}} - W_{o} - \chi
 - \mu}{kT}\right) + 1\right] ^{-2} dp.
\end{eqnarray}

   Denoting further
\begin{equation}\label{Sn}
S_{n} = \int_{0}^{\infty} \frac{p^{2}}{\exp \left( \frac{
 \sqrt{c^{2}p^{2} + W_{o}^{2}} - W_{o} - \chi - \mu}{kT}\right) + 1} dp,
\end{equation}
\begin{eqnarray}\label{SP}
S_{P} = \int_{0}^{\infty} \frac{p^{4}}{\sqrt{c^{2}p^{2} + W_{o}^{2}}}.
 \nonumber \\
 .\left[ \exp \left( \frac{\sqrt{c^{2}p^{2} + W_{o}^{2}} - W_{o} - \chi
 - \mu}{kT}\right) + 1\right]^{-1} dp,
\end{eqnarray}
and
\begin{eqnarray}\label{SE}
S_{E} = \int_{0}^{\infty} \frac{p^{2}}{\sqrt{c^{2}p^{2} + W_{o}^{2}}}.
 \nonumber \\
 .\left[ \exp \left( \frac{\sqrt{c^{2}p^{2} + W_{o}^{2}} - W_{o} - \chi
 - \mu}{kT}\right) + 1\right]^{-1} dp,
\end{eqnarray}
integrals (\ref{IP1}) and (\ref{IP2}) can be expressed (after one step of
the per partes integration) as
\begin{equation}\label{IP1b}
I_{P1} = 4kTS_{P} + \frac{3kTW_{o}^{2}}{c^{2}}S_{E},
\end{equation}
\begin{equation}\label{IP2b}
I_{P2} = \frac{3kT}{c^{2}}S_{n}.
\end{equation}
Since the integrals $S_{n}$, $S_{P}$, and $S_{E}$ can also be given as
$S_{n} = n_{gas}/K$, $S_{P} = 3P_{gas}/(Kc^{2})$, and
$S_{E} = (E_{gas} - 3P_{gas} + W_{o}n_{gas})/(KW_{o}^{2})$, the gradient
of pressure (\ref{gradP2}) can be expressed as
\begin{equation}\label{gradPfin}
\frac{dP_{gas}}{dr} = n_{gas}\frac{d\chi}{dr} + \frac{B}{T}\frac{dT}
 {dr}.
\end{equation}
In the last equation, we denoted
\begin{equation}\label{B}
B = P_{gas} + E_{gas} - \chi n_{gas} - \mu n_{gas}.
\end{equation}

\section{The solution for an extremely hot gas}\label{sec4}

   Let us consider the reference frame, to which the line element
(\ref{line_elem}) is referred. It means, it is the proper reference
frame of an infinitesimally small gas volume in the URMS surface layer
at the distance $r$ from the center of URMS. We further consider the
static case, i.e. the components of the metrique tensor
$g_{11} = \exp(\lambda)$ and $g_{44} = \exp(\nu)$ are the functions of
only the space coordinate $r$. According the outer Schwarzschild
solution of the Einstein's equations of field, it is well-known that
\begin{equation}\label{g11}
g_{11} = -\frac{1}{g_{44}} = -\frac{1}{1 - R_{g}/r}.
\end{equation}

   For $r \rightarrow R_{g}$, there appears the singularity
$|1/g_{44}| \rightarrow \infty$, which is essential in our deductions.
The behavior of the state quantities of gas are derived in the
coordinate system, in which this singularity appears. Its existence
implies that the conditions for the extremely high pressure and
temperature must exist not only at the center of the URMS (at
$r \rightarrow 0$, therefore $1/r \rightarrow \infty$), but around the
sphere of radius $R_{g}$ as well. This indicates that it is reasonable
to investigate, in the search for the URMS' stability, the properties of
the extremely hot, ultra-relativistic gas, the existence of which
appears to be very probable in the surface layer of the URMS with the
proper physical radius not much different from its gravitational radius.

   Close to $R_{g}$, we can obviously expect that
$\chi \rightarrow \infty$. So, let us search for the solution of
integrals (\ref{basicn})$-$(\ref{basicE}) assuming that the gas is
situated in the gravitational field characterized with the strong
inequality $\chi \gg W_{o}$. In other words, we neglect the rest energy
of the gas particles with respect to the magnitude of their relativistic
potential-type energy, $\chi$. Consequently, the relation (\ref{W})
reduces to
\begin{equation}\label{Wrel}
W = cp - \chi .
\end{equation}
As well, we neglect the chemical potential, $\mu$, because any chemistry
cannot be important in the considered extreme conditions.

   If both $W_{o}$ and $\mu$ are neglected, integrals
(\ref{basicn})$-$(\ref{basicE}) are simplified to
\begin{equation}\label{n2sol}
n_{gas} = K \int_{0}^{\infty} \frac{p^{2}}
 {\exp\left( \frac{cp - \chi}{kT}\right) + 1} dp,
\end{equation}
\begin{equation}\label{P2sol}
P_{gas} = \frac{Kc}{3} \int_{0}^{\infty} \frac{p^{3}}
 {\exp\left( \frac{cp - \chi}{kT}\right) + 1} dp,
\end{equation}
\begin{eqnarray}\label{E2sol}
E_{gas} = 3P_{gas},
\end{eqnarray}
respectively. One can notice that we also neglect $W_{o}$ in the
argument of square root $\sqrt{c^{2}p^{2} + W_{o}^{2}}$ in the
denominator of Eq.(\ref{basicP}), where the rest energy $W_{o}$ is
comparable to $cp$ at the down limit of this integration. Fortunately,
this neglection is possible, because the contribution of the integration
in this region of $p$ to the total integral is well-negligible.

   Eqs.(\ref{n2sol}) and (\ref{P2sol}) do not contain any particle
characteristics, except of constant of proportionality $K$. Since all
following relations are valid for every constant $K$, our considerations
concerning the hot gas can be generalized for an arbitrary
Fermi-Dirac-statistics gas consisting of several kinds of particles
characterized with different degeneracy states $f_{j}$ and having the
partial concentrations $n_{g,j}$. In the generalization, the constant
$K = 4\pi f_{1}$, for the $1/2$-spin mono-particle gas, must only be
replaced with constant $K = (4\pi/n_{gas})\sum_{j}n_{g,j}f_{j}$ for
the various-half-spin multi-particle gas.

   The integrals of type
\begin{equation}\label{genint}
I_{\gamma} = \int_{0}^{\infty} \frac{x^{\gamma}}{\exp (x - b) + 1}dx
\end{equation}
are well-known in quantum statistics. The approximative solution can be
found in form (e.g. \cite{madelung64})
\begin{eqnarray}\label{solInu}
I_{\gamma} = b^{\gamma + 1}\left[ 2 \sum_{j=1}^{\infty}
 \frac{\alpha_{2j}\Pi (\gamma )}{b^{2j} \Pi (\gamma - 2j + 1)} +
 \right. \nonumber \\
 \left. \frac{\Pi (\gamma )}{\Pi (\gamma + 1)}\right]
 + \Pi (\gamma ) \mathcal{R}(b,\gamma ),
\end{eqnarray}
where
\begin{eqnarray}
\alpha_{2j} = \sum_{l=0}^{\infty} \frac{(-1)^{l}}{(l + 1)^{2j}},
 \nonumber \\
|\mathcal{R}(b,\gamma )| \leq \exp(-b), \nonumber
\end{eqnarray}
and $\Pi$ is the factorial function. Below in this section we however
find the possibility that exponent $b$ in the condition delimiting
$|\mathcal{R}|$ may not be large, therefore $|\mathcal{R}|$ cannot be
neglected. For the values of $\gamma$ figuring in the considered
integrals (\ref{n2sol}) and (\ref{P2sol}), i.e. $\gamma = 2$ and 3, we
utilize the full results of integrations yielding
\begin{eqnarray}\label{n3sol}
n_{gas} = \frac{K\chi^{3}}{3c^{3}}\left[ 1 + \frac{12\alpha_{2}
 (kT)^{2}}{\chi^{2}} + \frac{6(kT)^{3}}{\chi^{3}} . \right. \nonumber \\
 \left. . \sum_{j=1}^{\infty}
 \frac{(-1)^{j+1}}{j^{3}}\exp\left( -j\frac{\chi}{kT}\right)
 \right] ,
\end{eqnarray}
\begin{eqnarray}\label{P3sol}
P_{gas} = \frac{K\chi^{4}}{12c^{3}}\left[ 1 + \frac{24\alpha_{2}(kT)^{2}}
 {\chi^{2}} + \frac{48\alpha_{4}(kT)^{4}}{\chi^{4}} - \right.
 \nonumber \\
 - \left. \frac{24(kT)^{4}}{\chi^{4}}\sum_{j=1}^{\infty}
 \frac{(-1)^{j+1}}{j^{4}}\exp\left(-j\frac{\chi}{kT}\right)
 \right] ,
\end{eqnarray}
where $\alpha_{2} = \pi^{2}/12 = 0.822467033$ and
$\alpha_{4} = 0.947032829$.

   Using Eq.(\ref{E2sol}) and neglecting $\mu$, the auxiliary quantity
$B$ (see Eq.(\ref{B})) is
\begin{eqnarray}\label{B2}
B = 4P_{gas} - \chi n_{gas}.
\end{eqnarray}
Using further Eq.(\ref{equivT}), we can find
\begin{equation}\label{gradTdivT}
\frac{1}{T}\frac{dT}{dr} = -\frac{1}{2}\frac{d\nu }{dr}.
\end{equation}
With the help of Eqs.(\ref{B2}) and (\ref{gradTdivT}), we can write the
relation (\ref{gradPfin}) in form
\begin{equation}\label{gradPqm}
\frac{dP_{gas}}{dr} = -2P_{gas}\frac{d\nu }{dr} + \left( \frac{d\chi }
 {dr} + \frac{1}{2}\chi \frac{d\nu }{dr}\right) n_{gas}.
\end{equation}

   For the strong field characterized with condition $\chi \gg W_{o}$,
function $\chi = \chi (\nu )$ can be found assuming that the URMS was
formed as a product of collapse of an object with so large radius, that
the magnitude of potential-type energy of gas particles was much lower
that their rest energy, i.e. $\chi \ll W_{o}$ for a single particle.
And, it was not extremely hot, therefore the kinetic energy of the
particles was also much lower than their rest energy, i.e.
$\sqrt{c^{2}p^{2} - W_{o}^{2}} - W_{o} \ll W_{o}$ for a single particle.
It means, whole energy of a particle consisted only of the rest energy
and we can, in the given context, well approximate the sum of kinetic
and potential-type energies by zero ($W_{k+p} = 0$) before the collapse.
The kinetic energy per one particle can also be given as
$E_{gas}/n_{gas}$. Since the total energy is expected to be conserved,
energy $W_{k+p}$ must be zero also after the collapse in the situation,
when $\chi \gg W_{o}$. This requires the validity of
$E_{gas}/n_{gas} - \chi = 0$.

   The energy density, $E_{gas}$, is proportional to the temperature.
Specifically, we express it as
\begin{equation}\label{EsnkT}
E_{gas} = sn_{gas}kT,
\end{equation}
where $s$ is a constant parameter. Using relation (\ref{EsnkT}) and fact
that $E_{gas}/n_{gas} - \chi = 0$, function $\chi$ is
\begin{equation}\label{zeroW}
\chi = skT.
\end{equation}
If we now use the Tolman \& Ehrenfest's relation (\ref{equivT}), we
obtain
\begin{equation}\label{chinuWo}
\chi = skT_{c}\exp(-\nu /2).
\end{equation}

   The constant $skT_{c}$ can be identified to $W_{o}$ because of the
following requirement. Let us consider the quantum-physics concept of
the wave associated with the particle. If the angular frequency of this
associated wave in the free space, far from the URMS, is $\omega_{o}$,
the wave-related energy of the particle is identical to its rest energy
and given by the well-known formula $W_{o} = \hbar \omega_{o}$, where
$\hbar$ is the Planck constant divided by $2\pi$. In the place of a
considered gas volume, where the potential-type energy is $\chi$, the
angular frequency is $\omega$ and the corresponding wave-related energy
of the particle is $W_{w} = \hbar \omega$. Energy $W_{w}$ is positive
and increases with the increasing magnitude of the gravity. It is
obviously a consequence of the fact that just this increase of energy
balances the corresponding decrease of the particle's potential-type
energy. Therefore, we can put $\chi = W_{w} - W_{o}$. Using further the
well-known transformation between $\omega$ and $\omega_{o}$, i.e.
$\omega = \omega_{o}/\sqrt{g_{44}}$, we finally find
\begin{equation}\label{Whomega}
W_{w} - W_{o} = \chi = \hbar \omega_{o}\exp(-\nu /2) - \hbar \omega_{o},
\end{equation}
or, after neglecting $W_{o}$,
\begin{equation}\label{chinu}
\chi = W_{o} \exp(-\nu /2),
\end{equation}
where we put $\hbar \omega_{o} = W_{o}$. Comparing the right-hand side
of this equation with the right-hand side of Eq.(\ref{chinuWo}), it is
clear that constant $skT_{c}$ must equal $W_{o}$.

   In the last step in the search for the URMS stability, let us look
for such a configuration of quantities that the quantum-statistics
solution for the gradient of pressure satisfies the equation of the
equilibrium (\ref{basicequ}). When Eq.(\ref{E2sol}) is also the valid
relation between the total internal energy density and total pressure,
i.e. it is valid that $E = 3P$ (this validity can be deduced from
Eqs.{\ref{E2sol}) and (\ref{E2solBE})), Eq.(\ref{basicequ}) changes to
\begin{equation}\label{gradPfield}
\frac{dP}{dr} = -2P\frac{d\nu}{dr}.
\end{equation}
We can see that this functional form for the gradient of pressure,
derived from the Einstein's field equations, is identical to that
derived within the quantum Fermi-Dirac statistics (see
Eq.(\ref{gradPqm})), if
\begin{equation}\label{source_chi}
\frac{d\chi }{dr} + \frac{1}{2}\chi \frac{d\nu }{dr} = 0.
\end{equation}
The validity of this condition can easily be shown differentiating
Eq.(\ref{chinu}), therefore Eq.(\ref{gradPqm}) becomes simplified to
\begin{equation}\label{gradPgas}
\frac{dP_{gas}}{dr} = -2P_{gas}\frac{d\nu}{dr}.
\end{equation}
If the radiation was completely absent, this would prove that the URMS
would be close to its stable configuration at whatever value of the
component $\exp(\nu )$ of metrique tensor or whatever radius $R_{b}$
for which the condition $\chi \gg W_{o}$ (or $W_{o}/\chi \ll 1$) is
satisfied. (The exact stability would appear for
$W_{o}/\chi \rightarrow 0$. A further discussion of the stability is
given in Sect.~\ref{sec6}.)

   The continuity of the gravity behavior implies that the function
$\exp(-\nu )$ as well as the derivative $d\nu /dr$ at the surface of
the URMS should approach their values calculated by the outer
Schwarzschild solution, according which
\begin{equation}\label{exp-nu}
\exp(\nu ) = 1 - \frac{R_{g}}{r}
\end{equation}
and, subsequently,
\begin{equation}\label{dnu}
\frac{d\nu}{dr} = \left( 1 - \frac{R_{g}}{r}\right) ^{-1}\,
 \frac{R_{g}}{r^{2}}.
\end{equation}
We can see that $d\nu /dr \rightarrow \infty$ for $r \rightarrow R_{g}$,
therefore also $dP_{gas}/dr \rightarrow \infty$ (see Eq.(\ref{gradPgas})
for $r \rightarrow R_{g}$.

   Integrating Eq.(\ref{gradPgas}), we can find
\begin{equation}\label{P_nu}
P_{gas} = P_{o}\exp(-2\nu ),
\end{equation}
where $P_{o}$ is an integration constant. This solution transparently
implies that $P_{gas} \rightarrow \infty$ when
$\exp(-\nu ) \rightarrow \infty$. Constant $P_{o}$ can be found
realizing the fact that the solutions (\ref{P3sol}) and (\ref{P_nu})
must be identical. Analysing these functional forms, we can find
\begin{eqnarray}\label{Po}
P_{o} = \frac{KW_{o}}{12c^{3}}\left[ 1 + \frac{24\alpha_{2}}{s^{2}} +
 \frac{48\alpha_{4}}{s^{4}} - \right. \nonumber \\
 \left. - \frac{24}{s^{4}}\sum_{j=1}^{infty}
 \frac{(-1)^{j+1}}{j^{4}}\exp(-js) \right] ,
\end{eqnarray}
where
\begin{equation}\label{s}
s = \frac{W_{o}}{kT_{c}} = const.
\end{equation}
with respect to the relations (\ref{equivT}) and (\ref{chinu}).


   For the cold, degenerated relativistic Fermi-Dirac gas, the relation
between the pressure, $P_{gas}$, and concentration of particles,
$n_{gas}$, was found in the form $P_{gas} \propto n_{gas}^{4/3}$ (see,
e.g., \cite{gursky76}). It is interesting that this proportionality is
also valid in the case of the extremely hot gas. From Eqs.(\ref{chinu})
and (\ref{P_nu}), we can obtain
\begin{equation}\label{P_chi}
P_{gas} = \frac{P_{o}}{W_{o}^{4}}\chi^{4}.
\end{equation}
Denoting further
\begin{eqnarray}\label{no}
n_{o} = \frac{K}{3c^{3}}\left[ 1 + \frac{12\alpha_{2}}{s^{2}} +
 \frac{6}{s^{3}}\sum_{j=1}^{\infty} \frac{(-1)^{j+1}}{j^{3}}.
 \right. \nonumber \\
 \left. .\exp(-js) \right] ,
\end{eqnarray}
we can re-write Eq.(\ref{n3sol}) as
\begin{equation}\label{n_chi}
n_{gas} = n_{o}\chi^{3},
\end{equation}
from which $\chi = (n_{gas}/n_{o})^{1/3}$. Supplying this into
Eq.(\ref{P_chi}), we obtain
\begin{equation}\label{P_n}
P_{gas} = \frac{P_{o}}{W_{o}^{4}n_{o}^{4/3}}n_{gas}^{4/3} =
 const. \times n_{gas}^{4/3}.
\end{equation}

\section{The influence of radiation}\label{sec5}

   Since the strong gravitational field of the URMS implies a high
internal energy of gas and, therefore, a high temperature, we can expect
an occurrence of an intensive radiation inside such a gas. The pressure
of this radiation has obviously to contribute to the total pressure. Let
us estimate an influence of this contribution on the stability of the
URMS, in this section.

   The statistical behavior of photons of electromagnetic radiation
can be described by the Bose-Einstein statistics. Because of a larger
context, we give a review of the well-known derivation of basic
relations for an arbitrary kind of bosons.

   In an analogy to Eqs.(\ref{n2sol})$-$(\ref{E2sol}) of the Fermi-Dirac
statistics, the concentration, pressure, and internal energy per unit of
volume can be written as
\begin{equation}\label{n2solBE}
n_{BE} = K \int_{0}^{\infty} \frac{p^{2}}
 {\exp\left( \frac{cp - \chi}{kT}\right) - 1} dp,
\end{equation}
\begin{equation}\label{P2solBE}
P_{BE} = \frac{Kc}{3} \int_{0}^{\infty} \frac{p^{3}}
 {\exp\left( \frac{cp - \chi}{kT}\right) - 1} dp,
\end{equation}
\begin{eqnarray}\label{E2solBE}
E_{BE} = Kc \int_{0}^{\infty} \frac{p^{3}}
 {\exp\left( \frac{cp - \chi}{kT}\right) - 1} dp = 3P_{BE},
\end{eqnarray}
for the photons, or other Bose-Einstein-statistics particles, when their
rest energy, $W_{o}$, is zero or can be neglected. Again, we inserted
the potential-type-energy term, $-\chi$, into the arguments of the
exponentials to completely characterize the energy state.

   Calculating integrals (\ref{n2solBE}) and (\ref{P2solBE}), we can find
\begin{eqnarray}\label{n3solBE}
n_{BE} = \frac{2K}{c^{3}}(kT)^{3}\sum_{j=1}^{\infty}
 \frac{1}{j^{3}}\exp\left(j\frac{\chi}{kT}\right) ,
\end{eqnarray}
\begin{eqnarray}\label{P3solBE}
P_{BE} = \frac{2K}{c^{3}}(kT)^{4}\sum_{j=1}^{\infty}
 \frac{1}{j^{4}}\exp\left(j\frac{\chi}{kT}\right) .
\end{eqnarray}
The solutions (\ref{n3solBE}) and (\ref{P3solBE}) for $n_{BE}$ and
$P_{BE}$ contain the sums, which are divergent for $\chi /(kT) > 0$. In
an expectation of a stable gas volume, this result implies that either
the same initial assumptions cannot be valid for both kinds of particle
statistics, Fermi-Dirac and Bose-Einstein, or $\chi /(kT) = 0$. It is
actually known that the bosons are not permanent constituents of matter.
In experiments, they are found to be unstable, except of photons,
therefore no law of the conservation of their number is satisfied, in
a common reality.

   For photons with zero rest mass and, thus, zero rest energy,
Eqs.(\ref{n2solBE})$-$(\ref{E2solBE}) are exact, not only approximative
equations. Consequently, the result of integrations expressed by
relations (\ref{n3solBE}) and (\ref{P3solBE}) is exact. Eq.(\ref{chinu})
implies that $\chi = 0$ for the photons, since $W_{o} = 0$ for them.
As mentioned above, the sums in relations (\ref{n3solBE}) and
(\ref{P3solBE}) are convergent for $\chi = 0$, therefore we have the
regular solutions for $n_{BE}$, $P_{BE}$, as well as $E_{BE}$
($E_{BE} = 3 P_{BE}$).

   For the radiation, the normalization constant $K_{N}$ is equal to
$K_{N}=1/(2\pi \hbar )^{3}$. The latter implies that
$K = 1/(\pi^{2}\hbar^{3})$. If $\chi = 0$, the pressure given by
Eq.(\ref{P3solBE}) becomes $P_{BE} \equiv P_{rad} = 2\beta_{4}(kT)^{4}/
(\pi^{2}\hbar^{3}c^{3}) = \pi^{2}(kT)^{4}/(45\hbar^{3}c^{3})$ (where
$\beta_{4} = \sum_{j=1}^{\infty}j^{-4} = \pi^{4}/90$). Or, with the help
of the radiation density constant, $a = \pi^{2}k^{4}/(15\hbar^{3}c^{3})$,
we can rewrite this relation to the form of another well-known formula
for the radiation pressure
\begin{equation}\label{Prad}
P_{rad} = \frac{1}{3}aT^{4}.
\end{equation}
The gradient of this pressure is
\begin{equation}\label{gradPrad}
\frac{dP_{rad}}{dr} = \frac{B_{rad}}{T}\frac{dT}{dr},
\end{equation}
where
\begin{equation}\label{Brad}
B_{rad} = \frac{4}{3}aT^{4}.
\end{equation}

   Using Eqs.(\ref{equivT}) and (\ref{Prad}), the quantum statistics
Eq.(\ref{gradPrad}) can be re-written to
\begin{equation}\label{gradPrad2}
\frac{dP_{rad}}{dr} = - 2P_{rad}\frac{d\nu }{dr},
\end{equation}
i.e. to the functional form identical to Eq.(\ref{gradPfield}) derived
from the Einstein's equations of field. As well, it is identical to the
functional form in Eq.(\ref{gradPgas}) valid for the gas consisting of
fermions. We see that the gradient of the radiation pressure behaves in
the same way as the gradient of the fermion-gas pressure.


\section{The last stage of the collapse above the event horizon}
\label{sec6}

   Since the individual gradients of both gas pressure and radiation
pressure acquire the values approaching those which are needed to
balance the gravity, as we demonstrated in Sects.~\ref{sec4} and
\ref{sec5}, the sum of these gradients also approaches that at which
these gradients would balance the gravitational attraction in the
extremely strong potential close to the Schwarzschild gravitational
radius. In terms of mathematics, the equation of the "asymptotic"
equilibrium between the total pressure gradient and gravity is the sum
of Eqs.(\ref{gradPgas}) and (\ref{gradPrad2}), i.e.
\begin{equation}\label{final_sum}
\frac{d(P_{gas} + P_{rad})}{dr} = - 2(P_{gas} + P_{rad})\frac{d\nu }
 {dr}.
\end{equation}
We emphasize that this-type requirement of the stable configuration of
URMS is set by Eq.(\ref{gradPfield}) yielding from the Einstein's
equations of field. Our proof of the balance (in the limit
$W_{o}/\chi \rightarrow 0$) is the fact that the same equation as
Eq.(\ref{gradPfield}) also occurs as the necessary description of
state quantities in the quantum statistics, which is represented by
Eq.(\ref{final_sum}). (We note $P_{gas} + P_{rad} = P$, in this
equation.) Both gravity and internal quantum state of the considered
gas/radiation converge to the same behavior close to the horizon, at
the gravitational radius.

  When applying the above conclusion about the balance in reality, we
must, however, keep in the mind that the condition of the balance is
$\chi \gg W_{o}$, or $W_{o}/\chi \ll 1$. The exact balance is set, when
$W_{o}/\chi \rightarrow 0$. This can be expected for
$r \rightarrow R_{g}$. For a larger $r$, we can expect a minute break
of the exact balance: in contrast to the before-URMS-stage fast collapse
(according to the theory of the last stages of stellar evolution), the
URMS either very slowly expands or collapses. If an expansion appeared,
it could be fuelled only by the internal energy of gas. Since there is
still a certain energy loss (see our considerations below), such the
expansion would earlier or later turn into a the collapse.

   In this section, we discuss some aspects of the last stage of the
assumed final, slow collapse of the URMS. We regard as the last that
stage of the collapse, when quantity $\exp(-\nu ) \gg 1$, but before the
moment when all the URMS' matter disappears under the horizon of events.
Unfortunately, we do not know the exact solution of general equations
(\ref{basicn}), (\ref{basicP}), and (\ref{basicE}) giving the quantities
of state as the functions of distance. Moreover, we would need a
solution of Einstein's equations of field for non-static case, which is
also unknown. A certain information can be deduced from a rough
comparison of relative ratios of the internal-energy loss of two URMSs
with different masses due to the escape of the energy to the outer
space.

   Considering the relation $E_{rad} = 3P_{rad}$ as well as
Eq.(\ref{Prad}), in which $T = T_{c}\exp{-\nu /2}$ with respect to
Eq.(\ref{equivT}), the energy of radiation in a unit surface volume,
$V_{1}$, is
\begin{equation}\label{EinV1}
W_{rad} = V_{1}aT_{c}^{4}\exp(-2\nu ).
\end{equation}
The energy irradiated through the unit surface area, $S_{1}$, of this
volume per a time unit can be given with help of the well-known
radiation law as
\begin{equation}\label{rad_flux}
\Phi = S_{1}\sigma T^{4} = S_{1}\sigma T_{c}^{4}\exp(-2\nu ),
\end{equation}
where $\sigma$ is the Stefan-Boltzmann constant.

   When we assume that the URMS' gas consists of neutrons with the rest
mass $m_{n}$, we can calculate $T_{c}$ using Eq.(\ref{s}), i.e.
$T_{c} = m_{n}c^{2}/(sk)$. In the thermodynamical equilibrium, the
temperature of gas and radiation is the same. In our ultra-relativistic
case, $s = 3$, therefore $T_{c} \approx 3.63\times 10^{12}\,$K.

   Eq.(\ref{rad_flux}) gives the radiation flux in the coordinate frame
of the emission source, i.e. in the frame of a volume in the URMS
surface layer. A distant observer detects the original radiation, having
the angular frequency $\omega$, with the red-shifted angular frequency
$\omega_{o}$, whereby
\begin{equation}\label{redshift}
\omega_{o} = \omega \exp(\nu /2).
\end{equation}
So, the flux measured by the distant observer is
\begin{equation}\label{rad_flux_red}
\Phi_{o} = S_{1}\sigma T_{c}^{4}\exp\left( -\frac{3}{2}\nu \right) .
\end{equation}

   The fraction $f_{rad}$ of radiation energy detected by the distant
observer per time $t$, which can be regarded as the loss of internal
energy of the surface volume $V_{1}$ per time $t$, is
\begin{equation}\label{f_rad}
f_{rad} = \frac{S_{1}\sigma}{V_{1}a}\, t \exp(\nu /2),
\end{equation}
in the simple case when $t$ is short enough that $f_{rad} \ll 1$. To
provide a specific numerical estimate of $f_{rad}$, we would need to
specify volume $V_{1}$. Unfortunately, it is impossible without an
explicit knowledge of functions $\lambda = \lambda (r)$ and
$\nu = \nu (r)$ for the URMS' interior. Nevertheless, we can expect the
continuous behavior of quantity $\nu$. At the URMS surface, it must
approach to that known from the outer Schwarzschild solution, i.e.
$\exp(\nu ) = 1 - R_{g}/R_{b}$. Therefore, $\exp(\nu /2) \rightarrow 0$
for $R_{b} \rightarrow R_{g}$. For a finite time $t$, the fraction of
the radiation-energy loss must also $f_{rad} \rightarrow 0$, when
$R_{b} \rightarrow R_{g}$. The URMS collapse can be expected to be
stopped at $R_{b} = R_{g}$.

   In a weak gravitational field with no significant red-shift, both
internal energy per a unit volume and irradiated energy from the volume
per a unit time are proportional to $T^{4}$, therefore the appropriate
$f_{rad}$ does not depend on temperature. There is no such the decrease
in the efficiency of cooling as in the above mentioned,
$\exp(\nu /2)$-proportional, ultra-relativistic case.

   We can expect a certain dependence of $f_{rad}$ and, thus, the
efficiency of cooling on the mass of the URMS. To compare, at least
very roughly, this dependence for the URMSs of two different masses, let
us consider a naive concept based on the assumption that the whole
internal energy of URMS is uniformly distributed in the volume
$4\pi R_{b}^{3}/3$. Considering further the corresponding surface
$4\pi R_{b}^{2}$ and identifying $R_{b} \doteq R_{g}$ (the difference
between $R_{b}$ and $R_{g}$ is a small fraction of $R_{g}$ during the
whole last stage of the URMS collapse), we can put
$S_{1}/V_{1} \approx 3/R_{g}$ implying
$f_{rad} \approx 3\sigma t \exp(\nu /2)/(aR_{g})$. Or, the time-scale,
on which fraction $f_{rad}$ of radiation energy is lost, can be
estimated as
\begin{equation}\label{t_loss}
t \approx \frac{aR_{g}f_{rad}}{3\sigma}\exp(-\nu /2) \approx
 \frac{2Gaf_{rad}}{3c^{2}\sigma} M \exp(-\nu /2).
\end{equation}

   The whole context of our study intuitively implies a replacement of
the old concept of the black hole by the concept of the URMS. Thus, not
only stellar-sized objects, but also the massive compact objects in the
galactic centers can be identified with the URMSs. The linear dependence
of the time-scale $t$ on mass $M$ given by Eq.(\ref{t_loss}) indicates
that the time-scale is about $10^{5}$ to $10^{10}$ times longer for a
super-massive URMS in a galactic center, with correspondingly larger
mass, than that for a stellar-sized URMS. This large factor indicates
that the collapse of the potential large URMSs in the universe cannot
obviously be the matter of a moment or a short period.\\

   Another potential mechanism of the internal energy loss is the escape
of neutrinos carrying energy. This mechanism is well-known to occur at
the first stage of the collapse of a massive stellar object. It is
verified observationally, so far, by the detection of neutrinos
preceding the optical outburst of supernova 1987A in the Large
Magellanic Cloud \cite{aglietta:etal89}.

   Although the extremely high pressure inside the URMS is keeping
the particle constituents of the gas, neutrons, stable, the neutrinos
can appear as a final result of collisions of photons, which are
abundant and very energetic in the URMS environment. Specifically, the
collision of two photons can produce an electron-positron pair, which
can again annihilate to produce the neutrino-antineutrino pair. Or,
neutrino (antineutrino) can occur in a collision of photon with a
lepton (occurred in the first reaction), which behaves as a catalysator.
Another reactions, including other elementary particles, can also occur
in a collision of photons. The important fact in our context are the
photons entering the reaction and neutrinos/antineutrinos as the result
of the reaction.

   The mean free path of neutrino in the gas with concentration of
screening particles $n_{gas}$ can be given as
\begin{equation}\label{neutr_path}
l_{\nu} \approx \frac{1}{\sigma_{\nu}n_{gas}}.
\end{equation}
If the energy of neutrino is $\mathcal{E}_{\nu}$, the cross-section
in the collision of a neutrino with a particle of the gas can be
estimated as \cite{hk94}
\begin{equation}\label{neutr_cs}
\sigma_{\nu} \approx \sigma_{o\nu}\mathcal{E}_{\nu}^{2},
\end{equation}
where the constant of proportionality, in SI units, was found to
equal $\sigma_{o\nu} \sim 10^{-11}\,$m$^{2}\,$J$^{-2}$.

   The neutrino cooling can only be sufficient if the energy carried
out by a single neutrino is comparable or higher than the internal
energy of gas per one gas particle (neutron), i.e.
$\mathcal{E}_{\nu} \goa E/n_{gas} = skT$. Supplying Eqs.(\ref{n_chi})
and (\ref{neutr_cs}) into Eq.(\ref{neutr_path}), whereby $\chi$ is
expressed using Eq.(\ref{chinu}), we easily find for the neutron gas
\begin{equation}\label{neutr_num}
l_{\nu} \loa \frac{1}{\sigma_{o\nu}n_{o}W_{o}^{5}}\exp(5\nu /2)
 \sim 10^{-16} \exp(5\nu /2)
\end{equation}
(in meters). Identifying constant $K$ to $K=1/(\pi^{2} \hbar^{3})$
in relation (\ref{no}), the value of constant $n_{o}$ is
$n_{o} = 2.253\times 10^{75}\,$m$^{-3}\,$J$^{-3}$. Since
$\exp(5\nu /2) \ll 1$, the mean free path of neutrino
$l_{\nu} \ll 10^{-16}\,$m. This path is much shorter than the scale
$R_{b} - R_{g}$ of the extreme-condition region, in which the reciprocal
metrique tensor $1/g_{44} \gg 1$. The neutrinos with a non-negligible
energy can escape only from the upper-most, extremely thin surface layer
together with the photons. Since they can originate only from photons,
an outflux of neutrinos cannot exceed the outflux of photons. The
appearance of neutrinos (by conversion from photons) only means that a
part energy, otherwise irradiated by photons, is carried out by
neutrinos.

   The efficiency of the neutrino cooling of the thin surface layer
correlates with that of radiation. The neutrinos also loose their
energy during their escape from the URMS surface due to the red-shift as
photons. And, either these particles cannot escape when the URMS radius
reaches the gravitational radius.

\section{Summary and conclusion}\label{sec7}

   In Sects.~\ref{sec4} and \ref{sec5}, we found the dependence of state
quantities on the auxiliary function $\nu$, for an extremely energetic
environment. This dependence on $\nu$ appears to be the same for both
particles satisfying the Fermi-Dirac statistics and photons satisfying
the Bose-Einstein statistics. In the case of the high energies, the rest
mass of fermions is negligible, therefore they seem to be as massless
particles as photons, and the identical dependence on $\nu$ is not very
surprising.

   In the beginning of Sect.~\ref{sec6}, we demonstrated that the
behavior of the gradient of total pressure, given by
Eq.(\ref{final_sum}), is identical to the relativistic equation of state
(\ref{gradPfield}), being a simplification of Eq.(\ref{basicequ}) in the
limit of infinite energy. In this point, we can see a unification of
quantum-physics description of state of ideal gas/radiation with the
general-relativistic description of space-time filled in with a gas and
radiation.

   Although we demonstrated that the efficiency of the URMS cooling is
low when the condition $\chi \gg W_{o}$ is valid, a certain energy still
escapes away and this energy loss obviously leads to shrinking of the
URMS until its proper physical radius becomes $R_{b} = R_{g}$. For
$R_{b} = R_{g}$, no regular energy loss is possible and the equilibrium
set by the identity of Eqs.(\ref{gradPfield}) and (\ref{final_sum}) is
exact. This equilibrium seems to be the law of Nature, supposed by
Eddington, which prevents very massive stars from the {\em reductio ad
absurdum}. According to our result, the final stage of an object with
a mass exceeding the Oppenheimer-Volkoff limit is the stable object
consisting of a matter and radiation and having radius exactly equal to
$R_{g}$. In other words, the outer border of the object is situated just
at the horizon of events.

   The concept of the space-time surrounding the URMS does not differ
from the classical concept of the space-time surrounding the
non-rotating, Schwarzschild black hole. Also the existence of the
horizon of events is predicted within both concepts. In our concept,
this horizon cannot be, however, regarded as an absolute, insuperable
barrier for any matter or radiation to escape. Just at the horizon,
there are permanently situated the gas and radiation having the internal
energy approaching infinity. In this context, it is necessary to note
that the derived equations are related to a perfectly quiet fluid with no
local phenomena. But in the reality, we do not know any perfectly quiet
object. Also on the URMS surface, some fluctuations of quantities of
state can be expected. Due to these fluctuations, some matter cannot be
excluded to occur above the horizon. If this happens, an emission of a
radiation can be expected. It must, of course, be extremely red-shifted,
but its eventual detection cannot be excluded. In this aspect, our
concept of the URMS is different to the concept of classical black
holes. (We started to use the term "URMS", instead of the traditional
"black hole", especially because of this reason.)

   To explain the basic principles of our new concept of the extremely
hot compact object more transparently, we constrained our considerations
to the simplest case of spherically symmetric, non-rotating object.
Despite this simplicity, we believe that the concept of URMS will be
helpful to explain some observational phenomena related to the
concerning objects.

\begin{acknowledgments}
   This work was supported, in part, by the VEGA, Slovak Grant Agency
for Science (grant No. 7047).
\end{acknowledgments}

\bibliography{phys_rev4}

\end{document}